\documentclass[12pt,preprint]{emulateapj}
\def\sref#1{\S\,\ref{sec:#1}}
\def\fref#1{Fig.~\ref{fig:#1}}
\def\tref#1{Table~\ref{tab:#1}}
\def\eref#1{Eq.~\ref{eq:#1}}
\defcitealias{duravk06a}{DvK06a}
\defcitealias{duravk06b}{DvK06b}
\newcommand\erf{{\rm erf}}
\begin{document}

\title{Extinction and Distance to Anomalous X-ray Pulsars from X-ray Scattering Halos}

\author{A. Rivera-Ingraham and M. H. van Kerkwijk}
\affil{Department of Astronomy \& Astrophysics, University of Toronto,
       50 Saint George Street, Toronto, ON M5S 3H4, Canada}
\email{rivera,mhvk@astro.utoronto.ca}

\begin{abstract}
  We analyze the X-ray scattering halos around three Galactic
  Anomalous X-ray Pulsars (AXPs) in order to constrain the distance
  and the optical extinction of each source.  We obtain surface
  brightness distributions from EPIC-pn data obtained with {\em
    XMM-Newton}, compare the profiles of different sources, and fit
  them with a model based on the standard theory of X-ray scattering
  by dust grains, both for a uniform distribution of dust along the
  line of sight, and for dust distributions constrained by previous
  measurements.  Somewhat surprisingly, we find that for all three
  sources, the uniform distribution reproduces the observed surface
  brightness as well as or better than the distributions that are
  informed by previous constraints.  Nevertheless, the inferred total
  dust columns are robust, and serve to confirm that previous
  measurements based on interstellar edges in high-resolution X-ray
  spectra and on modelling of broad-band X-ray spectra were reliable.
  Specifically, we find $A_V\simeq4$, 6, and 8\,mag for 4U~0142+61,
  1E~1048.1$-$5937, and 1RXS~J170849.0$-$400910, respectively.  For
  1E~1048.1$-$5937, this is well in excess of the extinction expected
  towards a \ion{H}{1} bubble along the line of sight, thus casting
  further doubt on the suggested association with the source.
\end{abstract}


\section{Introduction}
Among the diversity of objects comprising the neutron star family,
Anomalous X-ray Pulsars (AXPs) are one of the two classes of
magnetars, with their emission powered by the decay of extremely
strong magnetic fields, of $10^{14}$--$10^{16}\,$G, whose origin is
still a matter of debate (for a review, \citealt{woodt06}).  As
remnants of massive, short-lived stars, these sources lie
preferentially near the Galactic Plane and thus are strongly absorbed,
hindering measurements of their intrinsic spectra and luminosities.
In some cases, distances and extinction columns could be estimated
from associated supernova remnants (1E 2259+586,
\citealt*{gregf80,kothuy02}; 1E 1841$-$045, \citealt{vasig97}) and
other structures (1E 1048.1$-$5937, \citealt{gaen+05}).

More recently, \citet[hereafter
\citetalias{duravk06a},b]{duravk06a,duravk06b} tried to determine the
extinctions and distances to the sources more directly, by measuring
the extinction from absorption edges in high-resolution X-ray spectra,
and inferring distances by comparison with the run of extinction with
distance as inferred from red clump stars near the line of sight.
They found that all AXPs were located in spiral arms (not surprising
given their youth) and that their luminosities were remarkably
uniform, near $10^{35}{\rm\,erg\,s^{-1}}$.  Comparing the inferred
distances to earlier work, the results were consistent for 1E
1841$-$045 and XTE J1810$-$197 (for the latter a distance from
\ion{H}{1} absorption was available), but inconsistent for 1E 2259+586
and 1E 1048.1$-$5937.  For 1E 2259+586, \citetalias{duravk06a} argued
that the associated supernova remnant CTB 109 could be at the larger
distance of $\sim\!7.5\,$kpc instead of the $\sim\!3\,$kpc inferred by
\citet{kothuy02}, while for 1E 1048.1$-$45937, they concluded that, at
$\sim\!9\,$kpc, the source was not associated with the \ion{H}{1}
bubble at $\sim\!2.7\,$kpc found by \citet{gaen+05}, even though it
appeared fairly nicely centered on it.

Of course, it is possible that the analysis of
\citetalias{duravk06a},b was flawed.  The underlying assumptions are
that the extinction measures are reliable, that they reflect only
interstellar extinction, and that the run of interstellar extinction
with distance can be reliably inferred from red clump stars.  The
latter assumption is perhaps the safest, since the red clump stars are
fairly well-understood standard candles.  The extinction measures are
relatively uncertain, however, especially for 1E 1048.1$-$5937, which
relied on fitting of the broad-band X-ray spectrum and is thus
sensitive to the assumed intrinsic model (see \citetalias{duravk06b}).
Furthermore, if a large part of the extinction were local to the
source, and thus not present along the lines of sight to the red clump
stars, the distances would be overestimated.

Here, we seek to test the extinctions and distances using X-ray
scattering halos, which appear as diffuse emission around an X-ray
point source and arise from small-angle scattering of soft X-rays by
dust in the line of sight \citep{over65}.  X-ray halo analyses have
been carried out on a variety of galactic and extragalactic objects
\citep{maucg86,preds95}, and have proven to be particularly useful in
constraining the properties of dust grains in the ISM.  For variable
sources, where the time delays in the scattering halo could be
measured, they have also been used to determine distances
(\citealt{pred+00}, \citealt*{xianln07}).

For our purposes, the benefit of the scattering halo is mostly that it
provides a column density that is purely interstellar, since we can
only measure the halo outside the point-spread function (at
$\gtrsim\!0\farcm5$ for our {\em XMM-Newton} data), well beyond where
dust associated with the source would contribute.  Below, we will find
that the intensities of the halos confirm the extinction estimates of
\citetalias{duravk06b}, and that, therefore, there is little
extinction local to the source.  In principle, it should also be
possible to use the detailed structure of the halo to check the run of
extinction with distance inferred from the red-clump stars.  Here,
however, we find that, generally, the X-ray halos are fit just as
well, or even better, with dust distributions more uniform than
indicated by the data, perhaps because simplifications made in our
modelling of the dust removed some of the variety that would have
smoothed the structure further.

The layout of this paper is as follows.  In \sref{model}, we first
describe the model we use for fitting the X-ray observations.  Next,
in \sref{obs}, we describe how we measured the radial surface
brightness profiles from archival {\em XMM-Newton} data and, in
\sref{fits}, how we fit these with our model, discussing some
consistency checks and possible pitfalls.  We conclude in
\sref{conclusions}.

\section{Expected Surface Brightness Distributions}
\label{sec:model}

We model the observed surface brightness distributions for each source
with three components, an unscattered central point source, a halo of
light scattered by dust, and unrelated background emission.  Below, we
discuss our assumptions for the halo in more detail; for the point
source, we will use measurements of an unabsorbed source, and for the
background, we will assume it is independent of position.

The scattering of X rays on dust particles is, generally, a
complicated problem, and below we follow previous work in making a
number of simplifications, most of which are reasonable as long as one
considers X rays with energies above $\sim\!1\,$keV.  For our sources,
this also implies the single-scattering limit is appropriate.  

Assuming that the dust grains are spherical, one can write the
scattered intensity observed at an angular distance $\theta_{\rm obs}$
from the central source as \citep{smitd98},
\begin{eqnarray} 
I(\theta_{\rm obs})&=& F_{\rm X}
  \int^{E_{\rm max}}_{E_{\rm min}}S(E)dE
  \int^{a_{\rm max}}_{a_{\rm min}}N_{\rm H}n(a)da\nonumber\\
&\times& \int^1_0 \frac{\widetilde{f}(x)}{(1-x)^2} 
          \frac{d\sigma (E,a,\theta_{\rm scat})}{d\Omega} dx,
\label{eq:itheta}
\end{eqnarray}
where $F_{\rm X}$ and $S(E)$ are the observed flux and normalised
spectrum in the energy range $[E_{\rm min},E_{\rm max}]$; $a$ is the
grain radius and $n(a)$ the corresponding number density; $x$ is the
distance from the observer normalised by the distance to the source
(see, e.g., Fig.~1 in \citealt{mathl91}); $\widetilde{f}(x)$ is
the normalized spatial distribution of the scattering sites (equal to
1 if evenly distributed); and $d\sigma(E,a,\theta_{\rm scat})/d\Omega$
is the differential scattering cross section, with $\theta_{\rm scat}$
the angle over which the X ray is scattered.  For small angles,
$\theta_{\rm scat}=\theta_{\rm obs}/(1-x)$; hence, generally, the innermost
regions of the halo are due to dust grains close to the source, while
the outer regions are due to grains closer to the observer.

\begin{deluxetable*}{lllllll}
\tablewidth{0pt}
\tablecaption{XMM-Newton Imaging Data Used\label{tab:obs}}
\tablehead{
&&&\colhead{Exp.}&\colhead{$b^{\rm II}$}&\colhead{$A_{\rm V}$}&\colhead{$d$}\\
\colhead{Source}&\colhead{Obs.\ ID}&\colhead{Filter}&\colhead{(ks)}&\colhead{$(\arcdeg)$}&\colhead{(mag)}&\colhead{(kpc)}
}
\startdata
\multicolumn{4}{l}{\em Anomalous X-ray Pulsars}\\
4U 0142+61&\dataset[ADS/Sa.XMM#0112781101]{0112781101}&Thin&\phn4.2&\phn$-0.43$&$3.5\pm0.4$&$3.6\pm0.4$\\
1E 1048.1$-$5937&\dataset[ADS/Sa.XMM#0510010601]{0510010601}&Thin&32&\phn$-0.52$&$5.6:$&$9.0\pm1.7$\\
1RXS J170849.0$-$400910&\dataset[ADS/Sa.XMM#0148690101]{0148690101}&Medium&30&\phn$+0.03$&$7.7\pm2.2$&$3.8\pm0.5$\\[1.2ex]
\multicolumn{4}{l}{\em Empirical PSF}\\
PKS 2155-304&\dataset[ADS/Sa.XMM#0080940301]{0080940301}&Thin&39&$-52.25$&\nodata&\nodata\\
&\dataset[ADS/Sa.XMM#0158961301]{0158961301}&Medium&39&$-52.25$&\nodata&\nodata
\enddata
\tablecomments{All observations used EPIC-pn in the small window
  mode.  The exposure time is the total time not affected by flaring.
  For reference, the galactic latitude $b^{\rm II}$ is listed, as well
  as the extinction $A_{\rm V}$ and distance $d$ inferred by
  \citet{duravk06a,duravk06b}.  No error is listed for the extinction
  for 1E~1048.1$-$5937, since it is based on a fit to broad-band X-ray
  data and the uncertainty is thus dominated by the extent to which
  the assumed intrinsic emission model is correct.}
\end{deluxetable*}

In the above integral, the least-known parts are the dust size
distribution and the scattering cross section, both of which depend
on the physical state and composition of the dust grains.  For the
former, following \citet*{mathrn77}, generally a power-law form is
assumed, while for the latter either the Gaussian approximation of the
Rayleigh-Gans theory is used (hereafter RG; e.g., \citealt{over65}),
or a more detailed calculation is done using Mie theory
\citep{vdhu57}.  In the RG approximation, it is assumed that
reflection from the grain surface is negligible and that the phase of
the incident wave is not shifted inside the medium, resulting in
coherent addition of the scattered waves.  \citet{smitd98} have shown
that for $E\gtrsim1\,$keV and small scattering angles, the
approximation agrees well with Mie theory, while for lower energies, RG
theory underpredicts the scattering efficiency.

In our work, we will assume the RG approximation, and limit ourselves
to relatively high energies.  Specifically, we approximate the cross
section by \citep{vdhu57},
\begin{equation}
\frac{d\sigma(E,a,\theta_{\rm scat})}{d\Omega})\simeq
  C_{\rm dust}a^6\exp\left(-\frac{\theta_{\rm scat}^2}{2\widetilde{\sigma}^2}\right),
\end{equation}
where the constant of proportionality $C_{\rm dust}$ depends on dust
parameters such as the atomic charge, mass number, density, and
scattering factor \citep[e.g.,][]{henk+82}, and $\widetilde{\sigma}$
indicates the typical size of the halo produced by grains with radius
$a$, given by \citep{maucg86},
\begin{equation}
  \widetilde{\sigma}=10.4{\,\rm arcmin\;}
  \frac{1}{(E/1{\rm\,keV})(a/0.1{\rm\,\mu m})}.
\end{equation}

As an aside, we note that with this cross-section, and for
$\widetilde{f}(x)=1$, the integral over $x$ in \eref{itheta} can be
expressed in terms of the error function.  Furthermore, the integral
over annuli required to obtain mean surface brightnesses
around some $\theta_{\rm obs}$, can also be expressed easily.
Specifically, 
\begin{eqnarray}
\int_\theta\theta d\theta\lefteqn{\int_x \frac{1}{(1-x)^2}
  \exp\left(-\frac{1}{2}\left(\frac{\theta}{\widetilde\sigma(1-x)}\right)^2\right)dx}\nonumber\\
&=& \widetilde\sigma \theta\frac{\erf(\widetilde\alpha)}{\sqrt{2/\pi}}
  +\widetilde\sigma^2(1-x)\exp(-\widetilde\alpha^2),
\end{eqnarray}
where we defined
$\widetilde\alpha=\theta/\sqrt{2}\widetilde\sigma(1-x)$.  With this
expression, to calculate the expected halo brightnesses for piece-wise
uniform dust distributions $\widetilde f(x)$, we only need to
integrate numerically over $a$ and $E$.\footnote{For power-law dust
  distributions $n(a)\propto a^q$ with integral power $q$, the
  integral over $a$ can also be expressed in terms of error functions.}

For the dust size distribution, we use a power law $n(a)\propto a^q$
with two different choices for the power and lower and upper limits.
For our primary model, we set $q=-3.5$, following the finding of
\citet{mathrn77}, and use $a_{\rm min}=0.005{\rm\,\mu m}$ and
$a_{\rm max}=0.25{\rm\,\mu m}$ (informed by the work of, e.g.,
\citealt{consfp05}).  This model yields a halo dominated by somewhat
larger grains than found in the fits of \citet{preds95}, whose
calibration of scattering optical depth with optical extinction we
will use.  Therefore, to see how strongly our results depend on our
choice, we also try a somewhat steeper power, $q=-4$, and smaller
maximum grain size, $a_{\rm max}=0.18{\rm\,\mu m}$, more similar to
the values typically found by \citeauthor{preds95}.  Finally, we fit
profiles in narrow energy ranges, so that we can ignore the dependence
on the spectral shape~$S(E)$ in our integration over energy.

For given $\widetilde f(x)$, the above suffices to describe the radial
profile of the halo, and our fits yield the normalisation.
Integrating the shape over $\theta_{\rm obs}$ then yields the total
observed intensity of the halo, $I_{\rm halo}$, which is related to
the scattering optical depth $\tau_{\rm scat}$ by,
\begin{equation}
\frac{I_{\rm halo}}{I_{\rm halo}+I_{\rm core}} =
1-\mathrm{exp}(-\tau_{\rm scat}),
\end{equation}
where $I_{\rm core}$ is the unscattered intensity of the point
source.  

We normalize the scattering optical depth to 1\,keV using that
$\tau_{\rm scat}\propto E^{-2}$, and convert to equivalent visual
extinction using the empirical calibration\footnote{The relation given
  here supercedes the different numbers given in the text and abstract
  of \citet{preds95}; P.\ Predehl, 2008, pers.\ comm.}  derived from
29 Galactic sources by \citet{preds95},
\begin{equation} 
\tau_{\rm scat,1\,keV} = A_V(0.079\pm 0.003)-(0.052\pm0.019).
\end{equation}
We note that no offset would be expected, and its presence may reflect
the use of the RG approximation even at energies below 1\,keV, where
it is known to be inaccurate (P.\ Predehl, 2008, pers.\ comm.).  The
resulting systematic uncertainties are compensated to some extent,
however, by the fact that the assumptions we make are very similar to
those of \citet{preds95}, causing us to make errors in the same
direction.  Furthermore, assuming that the dust properties do not vary
wildly between different lines of sight, the same should hold between
our different sources, i.e., the relative intensities of the halos
should provide a good measure of the relative dust columns.

\section{Observed Surface Brightness Distributions}
\label{sec:obs}

In order to measure the scattering halos, we need good angular
resolution and sensitivity, yet avoid non-linearity in the bright,
unscattered light.  These constraints are best met by data taken with
the European Photon Imaging Camera with PN detectors (EPIC-PN;
\citealt{stru+01}) onboard {\em XMM-Newton} ({\em XMM}).  Most AXPs --
and most good reference sources -- are bright and thus observed with
small-window mode to avoid pile-up.  This helps to define the surface
brightness accurately close to the source, but also implies we cannot
measure the halo's full extent.  In the {\em XMM} archive, suitable
observations are available for five of the six well-studied anomalous
X-ray pulsars,\footnote{The sixth, XTE J1810$-$197, has only
  large-window data, for which there are no corresponding data of our
  point-spread function reference.  We tried to analyze it using the
  PSF derived from small-window data, but could not find a good model
  for the scattering halo.  This might be because of the different
  modes, the source's variability, or a break-down in our model.  We
  did not investigate this further.} but two of those (1E 1841-045 and
1E 2259+586) are associated with supernova remnants, the emission of
which makes it very difficult to measure the scattering halo.  Hence,
we limited ourselves to the three remaining sources, 1E 1048.1$-$5937,
4U 0142+61 and 1RXS J170849.0$-$400910; see \tref{obs}.  As a
reference source to define the point-spread function, we used the BL
Lac PKS 2155$-$304, which is at high galactic latitude and thus has a
negligibly small scattering halo, and which has been observed
regularly in different filters.

For all sources, we reprocessed the raw data with the standard
pipelines {\tt epchain} and {\tt emchain} from {\tt xmmsas} version
7.1.0 (and the calibration files available with that release).  We
obtained the surface brightness distributions by summing counts in
concentric annular regions of $10\arcsec$ width around the source,
with the innermost annulus starting at $6\farcs62$ (avoiding the
central parts, were systematic differences due to small centering
  errors or pile-up effects (see below) are expected to dominate) and
the outermost ending at $196\farcs62$.  We selected events within
given energy ranges using standard constraints for EPIC-PN (singles
and doubles, and no warning flags).  We corrected for out-of-time
(OoT) events by subtracting an appropriately scaled radial profile,
produced from simulated out-of-time only event lists (made with
  {\tt epchain}; we used the scale factor of 0.011 listed in the
  XMM-Newton Users Handbook for small-window mode).  We converted to
surface brightnesses by dividing by the areas, using the
\texttt{arfgen} and \texttt{backscale} tasks to ensure the areas
include only the parts of the annuli that overlap the detector.

A possible issue with our method, pointed out by the referee, is
  that the correction for out-of-time events might be unreliable if
  the source suffers from pile-up.  This is because our analysis
  relies on the fact that photons detected ``out of time'' -- during
  the transfer from the imaging part of the CCD to the storage area --
  have the same rate and spectrum as those measured during the
  integration.  If pile-up occurs, however, the measured source event
  rate is reduced and the measured spectrum hardened, and thus one
  would underestimate the rate of low-energy, out-of-time events.
  Fortunately, for our observations, all taken in small-window mode,
  pile-up is not an issue (for our brightest source, 4U 0142+61, the
  total count rate is $\sim\!50{\rm\,s^{-1}}$, while in small window
  mode, with its read-out rate of 175 frames per second, pile-up
  becomes significant only for rates in excess of
  $\sim\!100{\rm\,s^{-1}}$; we confirmed pile-up is not an issue using
  the {\tt epatplot} task).  Nevertheless, it would be a problem for
  brighter sources and/or imaging modes with longer integration times.
  In those cases, it would likely be best to measure the radial
  profile in areas excluding all parts affected by out-of-time events
  (i.e., excluding a strip along the read-out direction that passes
  over the source).

\section{Scattering Halos}
\label{sec:fits}

To derive halo intensities, we fit our radial profiles with a
combination of an empirical PSF, the scattering halo model for a
variety of assumed distributions of dust along the line of sight (and
for two dust size distributions; see \sref{model}), and a constant
background.  We fit the radial profile in two energy ranges
simultaneously, 1.2--1.6 and 2.0--2.4\,keV.  Here, the lower range
will have the stronger dust scattering signal (yet is at sufficiently
high energy that multiple scattering is not a significant issue for the sources considered here), but, given the small field of
view, it is mostly sensitive to dust close to the source.  The higher
energy range will have a weaker scattering halo (though still
noticeable), but helps to constrain the dust somewhat closer to the
observer, as well as to verify consistency.

We used the radial profiles from the observations of the high
  latitude source PKS~2155$-$304 as the empirical PSF in the fitting
  routines.  The main advantages over using a theoretical PSF are that
  one automatically minimizes any common artefacts (either
  instrumental or due to the analysis method; see, e.g., the wiggles
  at large radial distances in the right-hand panels of
  Figs~\ref{fig:4u0142halo}, \ref{fig:1e1048halo}, and
  \ref{fig:rxj1708halo}), artefacts not necessarily captured in a
  theoretical PSF model.  A disadvantage is that, effectively, the
  fits now measure relative brightnesses and background levels.  In
  order to put these on an absolute scale, we determined the
  backgrounds and normalizations for our calibrator by fitting its
  profiles using the theoretical PSF of \citet[][a King profile with
  energy and position-dependent width and exponent]{ghiz02}.

\begin{figure}
\begin{center}
\plotone{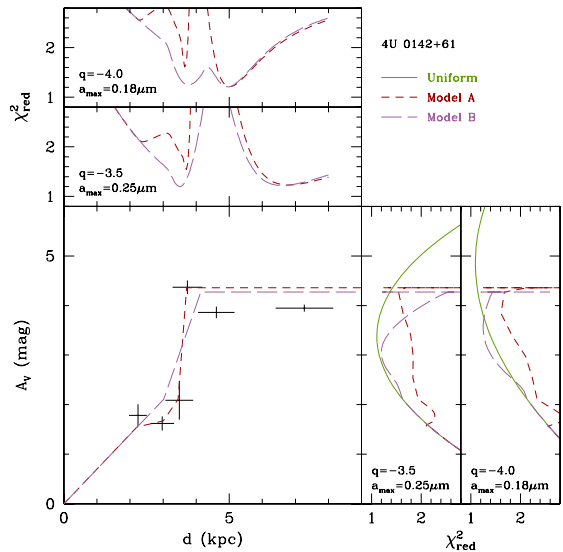}
\caption{Extinction and distance for 4U 0142+61.  In the main panel,
  the run of extinction with distance is shown, as inferred from
  red-clump stars by \citetalias{duravk06a}.  Overdrawn are two
  piece-wise uniform approximations we used to fit the halo intensity
  (models A and B).  In the top panels, we show for both models how
  the reduced $\chi^2$ of the fit to the observed radial profiles
  varies with distance, for two choices of dust distribution (as
  indicated).  In all cases, two minima appear, between 4 and 6\,kpc.
  In the right-hand panels, we show the implied variations in
  $\chi^2_{\rm red}$ with total extinction for both models and both
  dust distributions, as well as the results for a uniform dust
  distribution (for which the predicted scattering halos does not
  depend on distance).  For all models, one infers $A_V\simeq4$\,mag.}
\label{fig:4u0142dust}\label{fig:4u0142chi2}
\end{center}
\end{figure}

\begin{figure}
\begin{center}
\plotone{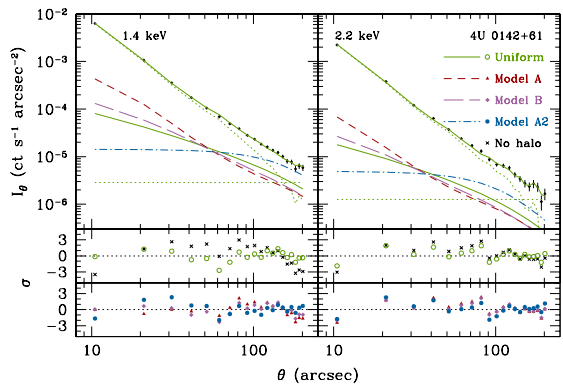}
\caption{Radial profiles of 4U 0142+61 in the 1.2--1.6\,keV ({\em
    left}) and 2.0--2.4\,keV ({\em right}) ranges, fitted with a
  combination of an empirical PSF, a scattering halo for a variety of
  models, and a constant background.  The green lines reflect the fit
  for a uniform distribution, with the dotted lines showing the
  unscattered and background components, and the solid lines the
  contribution from the halo and the total of all contributions.  The
  best-fit halo contributions from other models are also shown (as
  labelled).  In the two lower panels, deviations from the various
  fits are shown, with the residuals for fits using no halo or a halo
  for uniform dust in the middle panel, and those for fits for
  piece-wise linear dust distributions in the lowest panel.  Note that
  model A2 refers to the second minimum at large distance found for
  model A (see \fref{4u0142chi2}).}
\label{fig:4u0142halo}
\end{center}
\end{figure} 

For the dust distributions along the line of sight, we try two types.
First, we use a uniform distribution, in which the halo contribution
depends only on the total amount of dust.  Second, we try a variety of
piece-wise uniform distributions, constrained to follow the run of
reddening with distance inferred from red clump stars by
\citetalias{duravk06a}.  Here, the halo intensity depends on distance,
which determines both the total dust column, and the fractional
distances of the boundaries between the different pieces.  Since the
halo is fairly sensitive to the precise spatial distribution
of dust, we chose up to three different distributions.  For the first
-- model A -- we always tried to match the \citetalias{duravk06a}
results exactly (to the extent possible without having unphysical
  decreases of dust column with distance).  Since we found that sharp
changes provided worse fits, we also tried up to two further models --
B and C -- both chosen to be ``softer,'' yet still consistent with the
measurements of \citetalias{duravk06a}.  Combined, these models give a
good sense of the sensitivity of our results to such choices.

In total, any individual fit has five free parameters: two unscattered
source intensities and two background brightnesses (one for each of
the two energy ranges), and either the total dust column or distance
(which determines the fractional halo intensities).  In practice, in
our fits for uniform density, the five parameters are determined
simultaneously by linear least-squares, while for those with varying
dust distributions, we step through distance, and for each distance
calculate the expected fractional halo contribution and fit for the
corresponding best-fit unscattered source intensities and backgrounds.

\begin{figure}
\begin{center}
\plotone{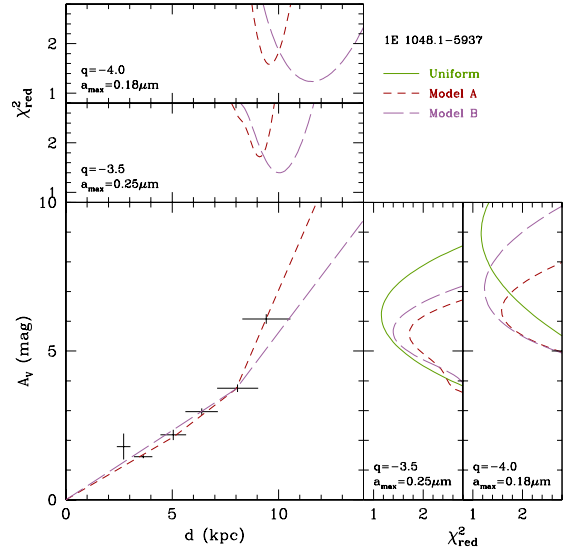}
\caption{Like \fref{4u0142dust}, but for 1E 1048.1$-$5937.  Note that
  both approximations to the run of reddening with distance use linear
  extrapolation beyond $A_V\simeq6\,$mag. The fits indicate a total
  extinction of about 6\,mag for our primary dust size distribution,
  and a distance of $\sim\!10\,$kpc.}
\label{fig:1e1048dust}\label{fig:1e1048chi2}
\end{center}
\end{figure}

We show our fit results in
Figs.~\ref{fig:4u0142chi2}--\ref{fig:rxj1708halo}, and summarize
numerical results for our primary dust distribution (with $q=-3.5$ and
$a_{\rm max}=0.25{\rm\,\mu m}$) in \tref{results}.  Two conclusions
stand out: (i) fits without a halo are much poorer than those with a
halo; and (ii) the fractional intensity of the halo increases as
expected from the reddening inferred from the X-ray spectra, with
4U~0142+61 having the smallest halo signal, 1E~1048.1$-$5937 an
intermediate one, and 1RXS J170849.0$-$400910 the largest.  Below, we
describe the results for the individual sources in detail, and then
discuss possible model caveats and systematic uncertainties.

\subsection{4U 0142+61} 

One of the brightest and most studied AXPs is 4U 0142+61.  Absorption
edges in its high-resolution X-ray spectrum indicate a total
extinction $A_{\rm V}=3.5\pm0.4$ \citepalias{duravk06b}, which,
combined with the run of extinction with distance inferred from the
red clump stars, suggests a distance of about 3.5\,kpc, in a region of
rapidly increasing optical extinction, presumably a spiral arm
(\citetalias{duravk06a}; see \fref{4u0142dust}).  Thus, we expect
that a large fraction of the dust along the line of sight is close to
the source, which should affect the inner parts of the halo in
particular.

In \fref{4u0142chi2}, we show how $\chi^2$ varies as a function of
reddening and distance.  For uniformly distributed dust with our
primary dust size distribution ($q=-3.5$ and $a_{\rm max}=0.25{\rm\,\mu
  m}$), there is a clear minimum around $A_V\simeq4.2{\rm\,mag}$,
close to the value inferred spectroscopically.  The piecewise uniform
models also show consistent minima, with reddening of 3--4\,mag and
distances of $\sim\!3.6$\,kpc.  However, they also have secondary
minima at larger distances, of $\sim\!6.7\,$kpc.  Given that AXPs are
young and thus likely to reside in spiral arms, the first minimum
almost certainly is the correct one.

For our second dust size distribution ($q=-4$ and $a_{\rm
  max}=0.18{\rm\,\mu m}$), we find fits of similar quality, but
somewhat larger values of the extinction.  This is as expected: for
smaller grains, more of the halo will be outside our field of view,
and hence a stronger halo is required to match the observed halo
contribution inside the field of view.  From these, we conclude the
uncertainty due to the dust distribution is considerable, although the
results do not change qualitatively.

In \fref{4u0142halo}, we show the fits to the radial profile, as well
as the residuals.  From the different halo curves, one sees how larger
quantities of dust concentrated towards the source increase the
innermost parts of the halo compared to the outer wings.  Indeed, both
piecewise models give very poor fits if the source is placed near the
far end or just behind the rapid rise of extinction around 3.7\,kpc
inferred from the red clump stars.  For the best fits, however, the
residuals are rather similar, since the changes in halo shape can be
compensated to some extent by changes in the other parameters, in
particular the background.  Note that in the higher energy range, the
halo is already very weak compared to the source: its fractional
intensity is about~5\%.

One surprise is that the uniform model gives a better fit than the two
piecewise uniform ones, even though the latter presumably are closer
to reality (as inferred from the red-clump stars).  Below, we find
that smoother distributions also fit better for the other sources.  We
briefly return to this in \sref{conclusions}, but stress here again
that among all models, the total extinction is fairly similar.

\subsection{1E 1048.1$-$5937}

One of the main goals of the present study was to test the extinction
and distance estimates for 1E 1048.1$-$5937.  Based on the coincidence
with a \ion{H}{1} bubble, Gaensler et al.\ suggested a distance of
$\sim\!2.7\,$kpc, but the high extinction measured for this source,
$A_{\rm V}\simeq6$, combined with the run of extinction with distance,
suggests a much larger distance of $\sim\!9\,$kpc \citepalias{duravk06a}. 

\begin{figure}
\begin{center}
\plotone{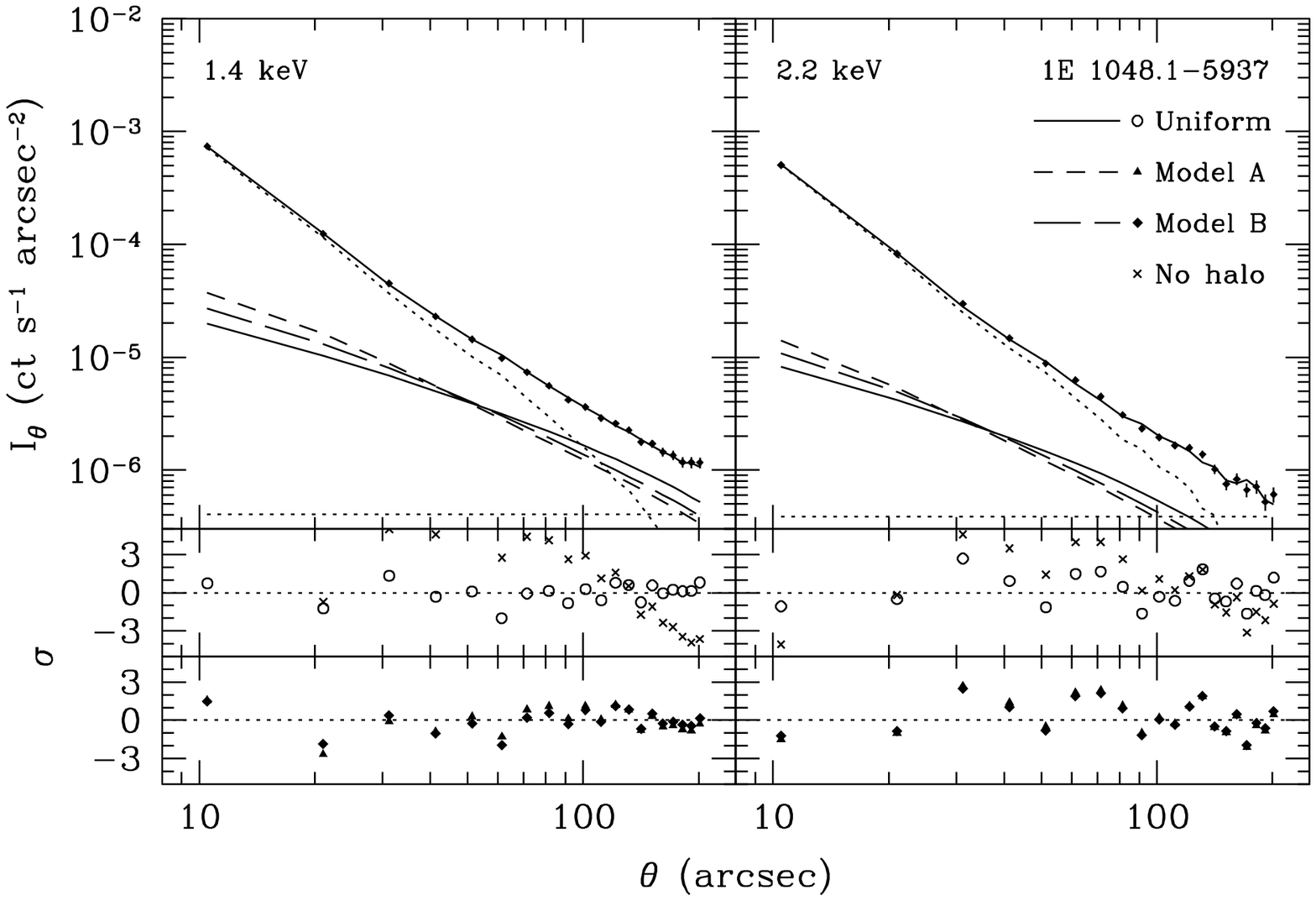}
\caption{Like \fref{4u0142halo}, but for 1E 1048.1$-$5937.}
\label{fig:1e1048halo}
\end{center}
\end{figure}

A major uncertainty is the estimate of the extinction, since it was
based on broad-band X-ray fits, and thus necessarily dependent on the
assumed underlying source model.  Unfortunately, the source was not
bright enough to detect absorption edges in its grating X-ray spectra
\citepalias{duravk06b}.  Visual inspection of observed radial profile
of 1E 1048.1$-$5937, however, already suggests the extinction is
indeed fairly high: the source clearly has a contribution due to a
scattering halo at least as large as 4U 0142+61 (compare, e.g., the
offset around 60\arcsec\ from the point-spread function in
Figs~\ref{fig:4u0142halo} and~\ref{fig:1e1048halo}).

This is confirmed by our scattering halo fits (\fref{1e1048chi2}): all
require $A_{\rm V}$ around 6\,mag (somewhat higher for our secondary
choice of dust distribution; note, though, that for those we had to
extrapolate the run of extinction with distance beyond the range
covered by the red clump stars).  Thus, like for 4U 0142+61, we find
that the scattering halos are consistent with the total extinction
inferred from the X-ray spectrum, and, based on the red-clump stars,
the inferred distance remains high, $\sim\!10\,$kpc.  

The high extinction (and strong halo) are inconsistent with the
extinction found from the red-clump stars at 2.7\,kpc.  The
explanation can also not be that there is strong extinction close to
the source, since this dust would contribute only to the very inner
parts of the halo, and thus not to the halo measured in our fit.

Comparing the different halo models in detail, we find again that the
fit for the more uniform dust distributions is the best.  Indeed,
following precisely the run of extinction with distance inferred by
\citetalias{duravk06a}, the fit is rather poor, with reduced
$\chi^2\simeq1.7$.  From the halos and residuals in
\fref{1e1048halo}, one again sees that the sharper increase in $A_{\rm
  V}$ near the source leads to a halo that is less consistent with the
observed one at small angles.

\subsection{1RXS J170849.0$-$400910}

Along the line of sight to 1RXS J170849.0$-$400910,
\citetalias{duravk06a} inferred large jumps in reddening from the red
clump stars (\fref{rxj1708dust}), likely associated with the Carina
and Crux spiral arms.  Given $A_{\rm V}=7.7\pm2.2$ inferred from
absorption edges in the X-ray spectrum \citepalias{duravk06b}, the
source should be inside the Crux spiral arm, at $3.8\pm0.5\,$kpc.

\begin{figure}
\begin{center}
\plotone{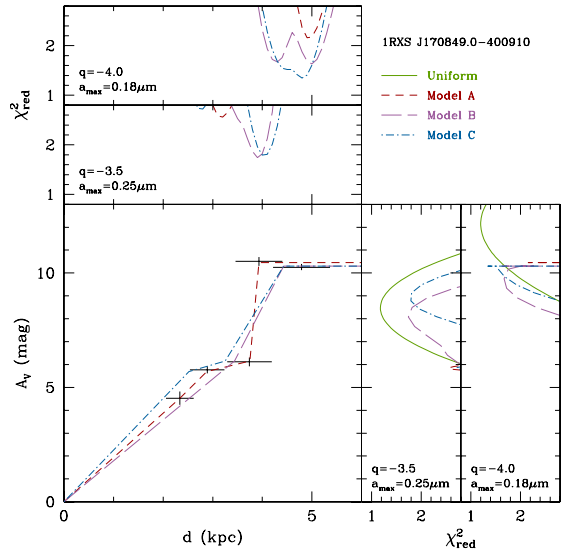}
\caption{Like \fref{4u0142dust}, but for 1RXS J170849.0$-$400910.  Here,
  three piece-wise uniform approximations are shown.  For model A,
  most fits have $\chi^2_{\rm red}$ values outside of the range shown
  in the top and right-hand panels.  From the other models for our
  primary dust size distribution, one infers $A_V\simeq8\,$mag and
  $d\simeq4\,$kpc.}
\label{fig:rxj1708dust}
\label{fig:rxj1708chi2}
\end{center}
\end{figure}

\begin{figure}
\begin{center}
\plotone{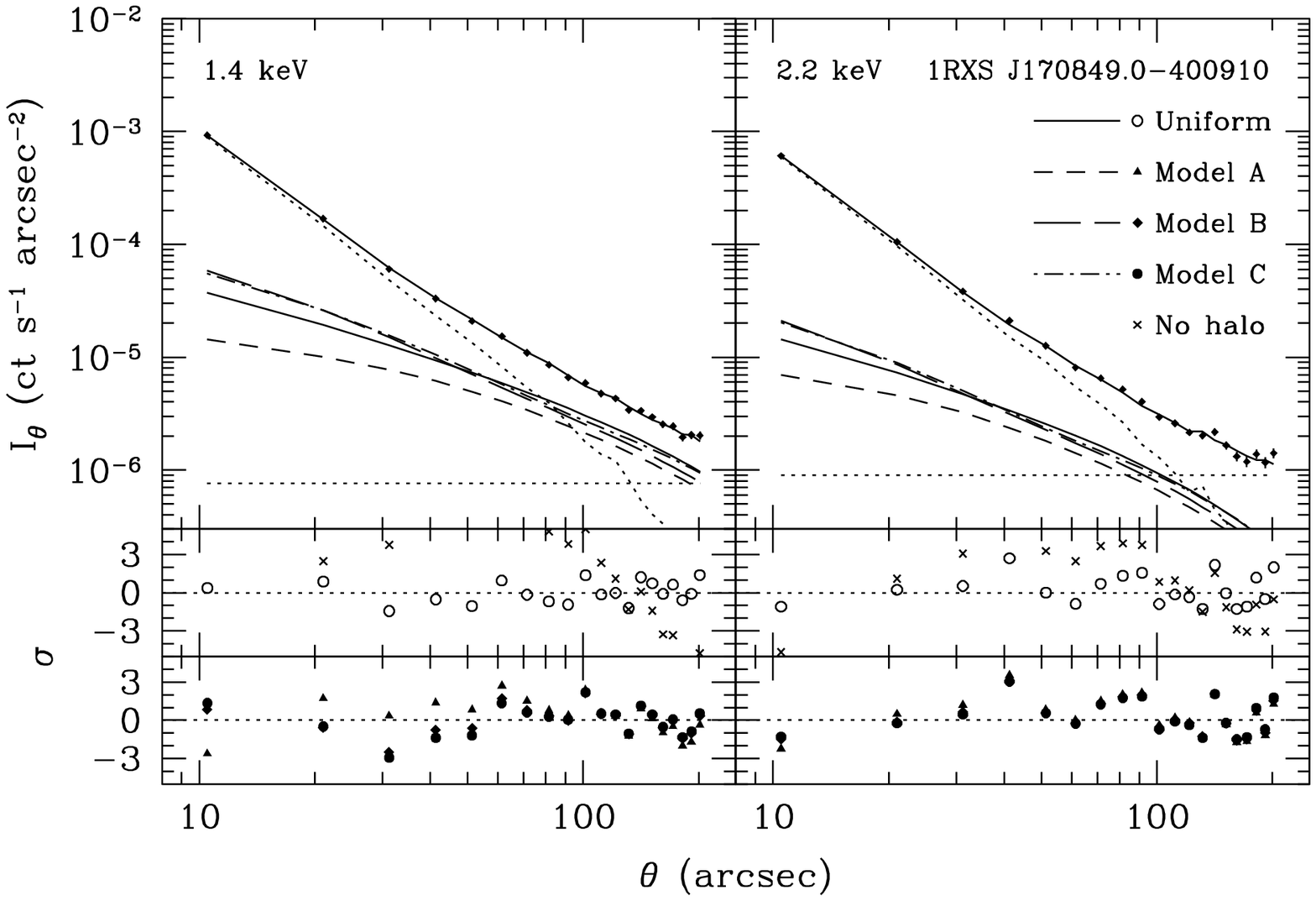}
\caption{Like \fref{4u0142halo}, but for 1RXS J170849.0$-$400910.}
\label{fig:rxj1708halo}
\end{center}
\end{figure}

Our fits generally reproduce the above, with $A_{\rm V}$ between 8 and
9\,mag, and the distance around $4\,$kpc.  Again, however, the fits
are much better for smoother dust distributions (\fref{rxj1708chi2},
\tref{results}).  Indeed, if we assume the exact dust distribution
inferred from the red clump stars, we obtain an unacceptable fit, with
reduced $\chi^2\simeq2.6$.  In \fref{rxj1708halo}, one sees that the
poor fit is due to the sharp increase in reddening near the source,
which causes a halo that is too strong and too flat at small angles
compared to what is required by the observations.

\begin{deluxetable*}{lrrrrrrr}
\tablewidth{0pt}
\tablecaption{X-ray Scattering Halo Fits\label{tab:results}}
\tablehead{&
\colhead{$f_{1.4}$}&\colhead{$f_{2.2}$}&
\colhead{$b_{1.4}$}&\colhead{$b_{2.2}$}&
\colhead{$A_{\rm V}$}&\colhead{$d$}&\\
\colhead{Model\tablenotemark{a}}&
\colhead{${\rm(s^{-1})}$}&\colhead{${\rm(s^{-1})}$}&
\colhead{${\rm(10^{-6})}$}&\colhead{${\rm(10^{-6})}$}&
\colhead{(mag)}&\colhead{(kpc)}&\colhead{$\chi^2_{\rm red}$\tablenotemark{b}}}
\startdata
\multicolumn{8}{l}{\em 4U 0142+61}\\ 
      No halo& 13.06&  4.68&  7.0&  1.9&\nodata&\nodata&  3.4\\ 
      Uniform& 12.59&  4.58&  2.9&  1.3&    3.4&\nodata&  1.1\\ 
           A & 11.87&  4.51&  4.7&  1.5&    4.2&    3.7&  1.5\\ 
           A2& 12.87&  4.60&  0.1&  0.7&    4.4&    6.8&  1.2\\ 
           B & 12.47&  4.57&  4.0&  1.4&    3.0&    3.5&  1.2\\ 
           B2& 12.87&  4.60&  0.3&  0.8&    4.3&    6.6&  1.2\\[1.2ex] 
\multicolumn{8}{l}{\em 1E 1048.1$-$5937}\\ 
      No halo&  1.55&  1.06&  1.5&  0.7&\nodata&\nodata&  9.5\\ 
      Uniform&  1.44&  1.02&  0.4&  0.4&    6.2&\nodata&  1.2\\ 
           A &  1.40&  1.01&  0.7&  0.5&    5.6&    9.1&  1.7\\ 
           B &  1.42&  1.01&  0.6&  0.5&    5.7&   10.0&  1.4\\[1.2ex] 
\multicolumn{8}{l}{\em 1RXS J170849.0$-$400910}\\ 
      No halo&  2.00&  1.31&  2.8&  1.5&\nodata&\nodata& 16.2\\ 
      Uniform&  1.83&  1.24&  0.8&  0.9&    8.5&\nodata&  1.2\\ 
           A &  1.91&  1.27&  1.3&  1.1&    5.8&    3.2&  2.6\\ 
           B &  1.78&  1.23&  1.1&  1.0&    8.1&    3.9&  1.7\\ 
           C &  1.78&  1.23&  0.9&  0.9&    8.9&    4.0&  1.8\\ 
\enddata
\tablecomments{These fits are for a dust size distribution $n(a)\propto
  a^{-3.5}$ between $a_{\rm min}=0.005$ and $a_{\rm max}=0.25{\rm\,\mu
    m}$.  We omitted formal uncertainties, since they are typically
  substantially smaller than the differences between different
  models.  The fluxes $f$ are those of the unscattered component; the
  background surface brightnesses $b$ have units of
  $10^{-6}{\rm\,s^{-1}\,arcsec^{-2}}$.}
\tablenotetext{a}{For the 4U~0142+61 models with a `2' suffix, the
  solution is that of a second minimum.  See \fref{4u0142halo}.}
\tablenotetext{b}{The number of degrees of freedom is 34 for the fits
  without halo, and 33 for fits with a halo model (38 data points, 4
  or 5 fit parameters).}
\end{deluxetable*}

\subsection{Background Consistency Check} 
\label{sec:background}

Within the small, $\sim\!256\times260\,$arcsec area covered by our
small-window mode observations, the profiles barely reach the
background, even without a scattering halo.  With a halo, which has a
typical size $2\widetilde\sigma\simeq1000\arcsec$ at 1.4\,keV, the
profile extends well outside.  Thus, one cannot determine the
background from an ``empty'' region, which is why we left it as a free
parameter.

Nevertheless, some consistency checks are possible.  Easiest perhaps
would be to compare with background estimates found in the outer
regions of EPIC-MOS exposures taken simultaneously with the PN
observations.  Unfortunately, we found that those ratios were not
constant, probably at least in part due to detector and position
dependent instrumental backgrounds \citep{cartr07}.  A rough sense can
be obtained from blank fields, even though no templates are available
for small-window mode.  These show that the typical backgrounds in the
two energy bands considered are both
$\sim\!0.5\times10^{-6}{\rm\,s^{-1}\,arcsec^{-2}}$, with at least
a factor 2 variation between different
observations.\footnote{\url{http://xmm2.esac.esa.int/external/xmm\_sw\_cal/background/index.shtml}}
These are roughly consistent with the rates we infer from our fits
(see \tref{results}).

A stronger constraint might be set by comparing the rates between
different energy bands.  This is because, as found by \citet{kata+04},
the spectrum of the background has no strong dependence on intensity,
at least at $\gtrsim\!1.7\,$keV.  Indeed, for our calibrator, we find
that we reproduce the ratio of $\sim\!1.1$ found by
\citeauthor{kata+04} between the background rate in the 4-7.2\,keV
range and that in the 1.7-4 keV range.  From our calibrator, we find
that for the bands we use, the ratio between the background counts
with those in the 4--7.2\,keV band are $\sim\!0.4$ and 0.2 (for
1.2--1.6 and 2.0--2.4\,keV, respectively; the ratio for the lower
energy band is more uncertain, consistent with the finding of
\citet{kata+04}, who found that the background rate below 1.7\,keV was
poorly correlated with the rate at higher energies).  We also find
that all observations have similar 4--7.2\,keV background count
rates, of 1.5--$2.5\times10^{-6}{\rm\,s^{-1}\,arcsec^{-2}}$.  Thus,
the expected background rates in the 1.2--1.6 and 2.0--2.4 bands are also
similar, $\sim\!0.8$ and
$0.4\times10^{-6}{\rm\,s^{-1}\,arcsec^{-2}}$.  This is roughly
consistent both to what we listed above for the blank fields, and with
what we infer from those of our fits that had acceptable~$\chi^2$.

Overall, we conclude that the fact that we cannot measure the
background rate directly, and thus have to leave it as a free
parameter in our fits, affects the precision of our analysis, but has
not introduced any major systematic uncertainties.

\subsection{Model Caveats \& Uncertainties} 

Our approach to fitting the halos has a number of potential pitfalls.
First, our observations cover only part of the scattering halo: within
the $\sim\!3\arcmin$ covered by the data, only about 60\% of the halo
signal is contained (for uniformly distributed dust).  However, to
first order, this should just lead to large uncertainties, not large
systematic errors: since we leave the background free, our limited
coverage simply means it is easier to obtain an adequate fit.
Furthermore, for our higher energy band, a bit more of the halo is
captured (about 75\% for uniformly distributed dust).

Second, the model we used to fit the observed scattering halo is
rather simplified and thus it is worthwhile to see what the possible
systematic errors could be.  One relatively small simplification is
that we take the mean energy of all our photons to be that of the
middle of a given energy band (and also use this to scale to the
scattering optical depth from that at 1\,keV).  The associated error
is likely small, however, since the count spectra of our sources show
that the photon energy distribution is quite uniform over our 0.4\,keV
bands.  (Note that we do integrate over the band for calculating the
scattering halo, so include the small amount of ``smearing'' because
of the change of energy.)

Another approximation we make is that we ignore multiple scattering.
\citet{mathl91} have shown that this dominates for $\tau_{\rm
  scat}>1.3$ or $A_V>17{\rm\,mag}$ at 1\,keV.  For our lowest energy
band, at 1.4\,keV, the use of the single-scattering approximation thus
requires $A_V\ll 33\,$mag, which is reasonably well satisfied for our sources.

The largest simplifications we make are the assumption of spherical
dust grains with a power-law size distribution, and the treatment of
scattering in the RG approximation.  For the dust distribution, our
two trial assumptions give significantly different results.  We tried
leaving the power-law index and limits free in our fits, but did not
reach a clear conclusion.  Generally, though, we would expect any
systematic errors should be in the same direction for all sources,
(unless dust properties vary significantly for different lines of
  sight, e.g., because one crosses a molecular cloud while another
  does not).  Thus, e.g., our conclusion that the total dust column
towards 1E~1048.1$-$5937 is about 50\% larger than that towards
4U~0142+61 should be reasonably safe.  Indeed, as already mentioned in
\sref{model}, since we use the calibration of the halo intensity with
extinction of \citet{preds95}, systematic errors may well be minimized
by using simplifications as similar to theirs as possible, as we do.

Despite the above, ideally one would use physical models for the
  dust, such as that proposed by \citet{weind01}.  Unfortunately, at
  present, our current understanding of dust scattering appears to be
  inadequate.  For instance, \citet{smit08} found that no model could
  reproduce in detail the dust scattering halo around GX 13+1, which
  he carefully measured in annuli ranging from 2 to 1000 arcsec.
  \citeauthor{smit08} also found that uniform dust distributions with
  power-law grain size distributions fitted the data as well as the
  presumably more physical models.  Thus, at present, it seems one
  will have to continue to rely on empirical calibrations between halo
  intensity and optical depth or reddening.

\section{Conclusions}
\label{sec:conclusions}

We used {\em XMM} observations to measure the strengths of the
scattering halos for three Anomalous X-ray Pulsars.  In all three
cases, we find that they are consistent with what was expected given
extinctions determined previously from fits to the X-ray spectra.
Combined with the run of extinction with distance, we thus also
confirm the distances found by \citetalias{duravk06a}.  In particular,
our results confirm that 1E 1048.1$-$5937 is much further than
inferred by \citet{gaen+05} from a possible association with an
\ion{H}{1} bubble.

Somewhat surprising is that the observed radial profiles are
consistently fit as well or better by uniform dust distributions than
they are by dust distributions that try to follow the actual run of
extinction with distance inferred from red-clump stars.  Part of the
reason for the poorer fits for the presumably more realistic dust
distributions may be that in our analysis we ignored complexities in
the dust properties or the scattering process that might tend to
smooth the radial profile, mimicking what would result from a more
uniform dust distribution along the line of sight.

Nevertheless, our results suggest that, if one calibrates one's
  analysis method with a source with known extinction (4U 0142+61 in
  our case), it is possible to infer estimates of extinction columns
  to other sources good to about 20\% from their scattering halos.
  With proper calibration, it should thus be possible to use the halos
  as independent constraints on model fits to the X-ray spectrum, and
  to help determine distances by comparing with the run of extinction
  with distance.

\acknowledgements We thank Martin Durant and Peter Martin for
discussions, the anonymous referee for a critical and constructive
reading, as well as the {\em XMM} help desk for answering a number
of queries.  This work is based on observations obtained with
XMM-Newton, an ESA science mission with instruments and contributions
directly funded by ESA Member States and the USA (NASA).

{\it Facilities:} \facility{XMM (EPIC)}

\end{document}